\documentclass[iop]{emulateapj}
\usepackage{color} 
\usepackage{ulem}

\def\delalpha{{\Delta \alpha \over \alpha}}

\def\kms{\rm km\,s^{-1}}
\def\ms{\rm m\,s^{-1}}
\def\lsim{{{}_{{}_<}^{~}\atop {}^{{}^\sim}_{~}}}

\def\ten#1{\times 10^{#1}}

\def\Nsys{N_{\rm sys}}
\def\Nmin{N_{\rm min}}
\def\Ntran{N_{\rm tran}}
\def\CNt{C_{\rm Nt}}
\def\lambdaI{\lambda_{\rm I}}

\def\lambdaTh{\lambda_{\rm ThAr}}

\def\vshift{v_{\rm shift}}

\usepackage{gensymb} %

\begin{document}

\title{Wavelength Calibration of the VLT-UVES Spectrograph}

\author{      
      Jonathan B.~Whitmore\altaffilmark{1},
      Michael T. Murphy\altaffilmark{2}, and
      Kim Griest\altaffilmark{1},
      \\
      }

\altaffiltext{1}{Department of Physics, University of California,
    San Diego, CA 92093, USA.}
\email{\tt jonathan.b.whitmore@gmail.com, mmurphy@swin.edu.au, kgriest@ucsd.edu}
\altaffiltext{2}{Centre for Astrophysics and Supercomputing, Swinburne University of Technology, Melbourne, Victoria 3122, Australia}

\begin{abstract} 
We attempt to measure possible miscalibration of the wavelength scale of the 
VLT-UVES spectrograph. We take spectra of QSO HE0515-4414 through the UVES 
iodine cell which contains thousands of well-calibrated iodine lines and 
compare these lines to the wavelength scale from the standard 
thorium-argon 
pipeline calibration.
Analyzing three exposures of this $z=1.71$ QSO, we find two distinct types of 
calibration shifts needed to correct the Th/Ar wavelength scale.  First, there is an overall average velocity
shift of between 100 $\ms$ and 500 $\ms$ depending upon the exposure.  
Second, within 
a given exposure,
 we find intra-order velocity distortions of 100 $\ms$ 
up to more than 200 $\ms$.
These calibration errors
are similar to, but smaller than, those found earlier in 
the Keck HIRES spectrometer.
We discuss the possible origins of these two types of miscalibration.
We also explore the implications of these calibration errors 
on the systematic error in measurements of $\delalpha$, the change in the
fine-structure constant derived from measurement of the relative
redshifts of absorption lines in QSO absorption systems.  
The overall average, exposure-dependent shifts should be less
relevant for fine-structure work, but the intra-order
shifts have the potential to affect these results.
Using either our measured calibration offsets or a Gaussian model with
sigma of around 90 $\ms$, Monte Carlo mock experiments find
errors in $\delalpha$ of between $1\ten{-6} \Nsys^{-1/2}$ and $3\ten{-6} 
\Nsys^{-1/2}$, where $\Nsys$ is the number of systems used and the range
is due to dependence on how many metallic absorption lines
in each system are compared.
\bigskip
\end{abstract}

\vskip .2truein
\bigskip

\section{Introduction}
In a recent paper (Griest et al. 2010) it was shown that Keck 
HIRES spectrograph measurements of absorption features in high redshift QSOs 
calibrated with the normal thorium-argon (Th/Ar) technique
contained overall average velocity shifts between exposures of up 
to 2000 $\ms$ over several 
nights and calibration errors of up to 500 $\ms$ even within a single 
echelle order in a single exposure.
Similar wavelength calibration problems were found previously by 
Osterbrock et al. (2000) and Suzuki et al. (2003).
These results were found by re-calibrating the wavelength scale using spectra
taken through the Keck HIRES iodine cell.  The iodine cell, which has been
used extensively in the Doppler method search for extrasolar planets,
puts several sharp lines per Angstrom on top of the QSO spectrum, and 
its reproducible absorption spectrum allows extrasolar 
planet 
workers to attain a relative velocity precision of around 5 $\ms$
between exposures taken up to several years apart (Johnson et al. 2006;
Konacki 2009).  
The analysis in Griest et al. (2010) compared the iodine
cell wavelength calibration to Th/Ar calibration with two separate Keck HIRES
analysis pipelines (XIDL and MAKEE) and so is probably measuring 
calibration problems with
the Th/Ar calibration of the spectrograph itself rather than problems 
just with the standard calibration software.

For most astronomical work, wavelength calibration errors of around 1 $\kms$ are not
important, but over the past few years several groups have used absorption lines in 
the spectra of high redshift QSOs to measure or place limits on possible changes
in the fine-structure constant over cosmological time.  
There has been considerable controversy in 
experimental measurements of $\delalpha$ using high redshift absorption 
systems, with claims of detection and also limits on variation inconsistent 
with those claims.  See Murphy et al. (2008) for a summary.

If the fine-structure constant was different 10 billion years ago by 
the claimed 
detection of $\delalpha = (-5.7 \pm 1.1)\ten{-6}$
(Murphy et al. 2003, 2004), 
relative atomic transition wavelengths are 
expected to differ from their lab values by up to $\sim$200 $\ms$.
For such experimental
uses, it is important to pay careful attention to calibration errors of the size
reported above.  However, it is important to understand that while 
individual measurement errors may be larger than the expected signal,
if many atomic transitions are used in many QSO absorption systems,
it may be possible to average away these calibration uncertainties.
See, for example, Murphy et al. (2009). 

The results of Griest et al. (2010) apply only to the Keck HIRES instrument,
but there is another instrument, the VLT UVES spectrograph, that is playing
a key role in the search for possible changes in the fine-structure constant 
using absorption lines in high redshift QSOs.
In this paper, we perform a similar recalibration of the standard UVES Th/Ar
wavelength calibration pipeline using the VLT iodine cell.  
We find similar, but smaller,
wavelength calibration errors than found in HIRES.  
We discuss the possible origin of these offsets, in particular
whether they arise from the UVES pipeline software or systematic errors
within the telescope and/or spectrograph.
We also make a first attempt at calculating
whether these calibration errors can give rise to important systematic errors
in the measurements to date of $\delalpha$.

\section{Observations}

\begin{deluxetable}{ccc}
\tablecaption{Journal of Observations
\label{tab:journal}
}

\tablewidth{0pt}
\tablehead{
\colhead{Exposure and Date}  & 
\colhead{Time (UT)} & 
\colhead{Th/Ar Time (UT)}
}
\startdata
1 2003 Oct 11 & 07:32 & 10:56 \\
2 2003 Oct 11 & 08:13 & 10:56 \\
3 2003 Oct 13 & 05:31 & 07:06 
\enddata
\end{deluxetable}

Six exposures were taken of the quasar HE0515-4414 ($z=1.71$, $V\approx14.9$\,mag) with the VLT-UVES 
spectrograph in 2003 October. 
In this paper, we analyze the wavelength calibration of the three exposures that 
were taken with the iodine cell in place.  
We include a journal of these observations in Table~\ref{tab:journal}.
Over the wavelength range of interest, the median signal/noise of the spectra
extracted from these exposures is around 20 pixel$^{-1}$ for the upper ``u'' chip and around 11 pixel$^{-1}$
for the lower ``l'' chip.  One pixel corresponds to about 1.5 $\kms$ at the leading edge of each
echelle order and around 0.9 $\kms$ at the trailing edge.
We note that ESO's specifications for UVES are that the gratings, after
being moved, 
be returned to the same position to within a tolerance 
corresponding to 0.1 pixels
(D'Odorico et al. 2000).
Thus naively we might expect to see an overall non-zero velocity calibration
shift between the iodine and Th/Ar lines of roughly 140 $\ms$ at 5500 \AA. 
The six QSO exposures were taken during two nights, with the first 
two I$_2$ QSO exposures being taken on the first night and the third I$_2$ exposure
taken on the next.  The first two exposures were calibrated 
with the same Th/Ar exposure;  however we note that
the gratings were moved after the two data exposures and then moved back to
the same position in order to take the Th/Ar exposure.
The third I$_2$ QSO exposure was followed first by a non-I$_2$ QSO exposure 
(same QSO, grating setting, etc.) and then by 
the Th/Ar calibration exposure.  We had hoped this scheduling
would remove a possible source of error caused by grating movement, 
but in fact, the third QSO I$_2$ exposure was part of a different
``observation block'', meaning the gratings were
reset (moved and then returned) between the data and Th/Ar exposures.
We therefore expect overall velocity shifts between the I$_2$ and Th/Ar wavelength
scales of order 140 $\ms$ for all three of our QSO exposures through the I$_2$ cell.

We can also estimate an overall average velocity shift having to do with the position of the QSO 
in the spectrograph slit.  The slit width is 0.7'' and our exposures were taken with seeing 
between 0.65'' and 0.85''.  This slit-width projects onto the CCD with an FWHM of about
4.8 $\kms$ ($R \sim 62,000$).  Therefore, if a given exposure has, for example, a 0.1'' positioning
error, we might expect a roughly 600 $\ms$ overall calibration shift, substantially larger than
the error caused by the resetting of the spectrometer grating.

In more detail, we note that each 2400 s I$_2$ exposure used only the red
arm of UVES in the standard 600 nm central wavelength setting for
I$_2$ observations. The red arm of UVES has a detector containing
two CCD detectors covering the wavelength ranges 496--597 and 599--707\,nm.
No on-chip binning was used; the pixels have a width
of $\approx1.3 \kms$, providing $\approx$3.7 pixels per FWHM resolution element.
A circular baffle, or ``pupil stop'', is used routinely in UVES to provide a beam from 
calibration lamps (e.g.,~Th/Ar, flat field) similar in size to that from the telescope from 
an astronomical point source. Our I$_2$ exposures were taken
with an under-sized pupil stop, i.e.,~a slightly smaller beam than
usual was allowed into UVES, but our Th/Ar exposures used a slightly
over-sized pupil stop.  While these are the default settings for
I$_2$ observations, strictly speaking our aim here is to treat the
I$_2$ and Th/Ar exposures as similarly as possible so that any
wavelength shifts between the two are appropriate to normal QSO
observations where a slightly ``oversize'' pupil stop is used for both
object and calibration exposures. However, we would not expect
slight, circular vignetting/truncation of the beam to
affect our results here. Indeed, subsequent iodine-cell tests with
UVES have shown any effect from the under-sized pupil stop to be
very small, if present at all; these results will be presented in a
forthcoming paper.

\section{Data Reduction and Analysis}
The QSO flux was extracted using the standard pipeline recipes. 
Five bias and flat field exposures were median filtered to
produce master bias and flat field corrections. The echelle order
positions and overall spectrograph setup were derived from
short-and-narrow slit exposures of quartz and Th/Ar lamps which were
used to refine a physical model for the expected flux distribution
within each order. In our exposures, the QSO flux had high enough
signal/noise in each echelle order to allow the flux distribution itself
to define its spatial profile for use as object weights in the
subsequent optimal extraction. No redispersion of the spectra was performed
after optimal extraction; each extracted echelle order retained its
original wavelength dispersion as a function of pixel
position. Rather than co-adding extracted QSO spectra, we treated
each order of each exposure separately so that each I$_2$ (and
accompanying Th/Ar) exposure gives a separate measurement of the
Th/Ar wavelength calibration shifts for each order.

The UVES Common Pipeline Language (CPL) software package 
includes considerable improvements to
the wavelength calibration process compared to previous UVES
reduction pipelines. The Th/Ar line list is only a small subset of
all known Th/Ar lines in the relevant wavelength range and was
derived via an objective line selection algorithm detailed in Murphy
et al. (2007).  An even smaller subset is used in the final
calibration, after the CPL pipeline removes those appearing too weak
or strong/saturated in the individual Th/Ar spectra.  The Th/Ar
exposures were bias- and (normalized) flat field corrected during
the extraction. The extractions used the spatial weighting profile
derived from the optimal extraction of the QSO flux from the
corresponding I$_2$ exposure. This ensures that the same pixels,
with the same weighting, contribute to both the QSO and Th/Ar
spectrum and naturally avoid calibration errors from small tilts of the
Th/Ar lines with respect to the CCD grid. A crude blaze correction
was made using the flat field spectral shape in each echelle
order. To determine their centroid in the wavelength calibration
process, the CPL pipeline fits Gaussian functions to the selected
Th/Ar emission lines. The Gaussian model includes an underlying
linear (rather than just constant offset) continuum level; this
reduces centroiding errors induced by other nearby emission lines or
other sources of background slopes, e.g., residual blaze function
(Murphy et al. 2007).

The wavelength calibration residuals around the default two-dimensional fourth-order polynomial wavelength solution had an rms of
$\sim30$--$40$ $\ms$, similar to that found from the unbinned
Th/Ar exposures by Murphy et al. (2007).  
We also performed wavelength
calibration using higher order polynomials and we study the effect of
this in Section 4.1.

To perform the wavelength recalibration using the iodine lines, we used a method
similar to that used in Griest et al. (2010).  A well measured iodine 
cell absorption spectrum taken elsewhere
is convolved with a Gaussian to give it the same resolution as the iodine lines
in our QSO spectra. It is then multiplied by an overall normalization
and shifted in wavelength by an amount that gives the smallest 
possible $\chi^2$ in
the difference between the convolved iodine spectrum and the QSO spectra.
Care is taken in the continua fitting as discussed in Griest et al. (2010),
since the iodine lines cover nearly the entire spectrum.  For the
reference iodine
spectra we tried both the Marcy and Butler (Butler, et al., 1996; Marcy 2008, unpublished)
Fourier Transform Spectrograph (FTS) spectrum of the Keck HIRES iodine cell done at KPNO with a 
resolving power, $R=170,000$
and a signal/noise of 700 pixel$^{-1}$, and the UVES iodine cell calibration 
spectrum,\footnote{http://www.eso.org/sci/facilities/paranal/instruments/uves/tools}
performed at 70\degree C, with a spectral resolution 0.020 cm$^{-1}$ 
(which implies $R>1,000,000$ throughout the effective I$_2$ wavelength range).
Besides the difference in resolution and a single shift in the absolute scale, 
the resulting calibration shifts were fairly similar using the two 
different reference spectra.
For UVES iodine cell data, it is clearly
more appropriate to use the UVES FTS iodine cell spectrum 
so we will only present results obtained using this spectrum.

\section{Results}
The result of our analysis is a wavelength-dependent velocity shift $\vshift(\lambda)$ between
the iodine cell value, which is presumed to be correct, and the Th/Ar value output by
the UVES pipeline software,
\begin{equation}
\lambdaI(\lambda) = \lambdaTh(\lambda) +\vshift(\lambda),
\label{eqn:io}
\end{equation}
where there is one such velocity shift function for each echelle order of each exposure.
We use the iodine cell calibration as the standard since it has a much
higher density of lines and because the QSO and the iodine light follow the
same optical path, have the same instrument illumination, and are
simultaneous.  
Additional evidence in support of the superiority of the iodine cell wavelength
scale and discussion of whether $\vshift$ arises primarily from the extraction software or from
optical distortion inside the spectrograph are given below.
In fact, it is not important to our work that the FTS iodine spectrum be
absolutely correct, since we only care about relative shifts across an
individual exposure, and perhaps shifts over time.  
And since the extraction of the Th/Ar spectra was performed using the same object weights as 
that of the QSO flux, the I$_2$ and Th/Ar spectra are treated as similarly as possible 
in the reduction process, meaning that any wavelength distortions derived should be driven 
by optical distortions rather than any deficiencies of the reduction pipeline.

\begin{figure}
    \includegraphics[angle=-90,width=0.5\textwidth]{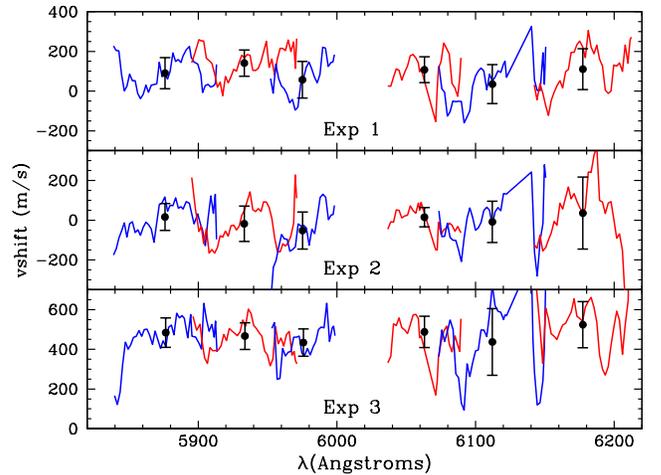}
    \caption{
    Shift in $\ms$ needed to bring the Th/Ar VLT UVES calibration in line with 
    the iodine spectrum.  
    This figure shows results for the three exposures orders of the UVES upper ``u'' CCD chip, with blue and red 
    colors alternating for each echelle order.  
    The black dots show the weighted average of
    the shifts in each order, with error bars showing the 
    average 1$\sigma$ fit error for each order. }
    \label{fig:vshiftallu}
\end{figure}

\begin{figure}
  \includegraphics[angle=-90,width=0.5\textwidth]{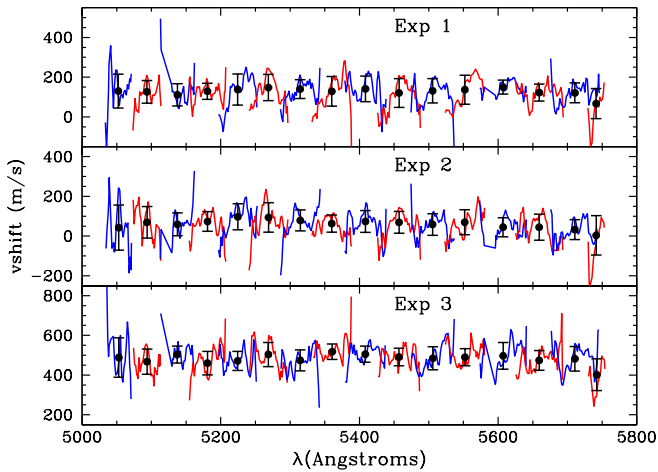}
  \caption{Same as the previous figure, but for the lower ``l'' CCD chip.
  \label{fig:vshiftalll}
  }
\end{figure}

The fit for the $\vshift$ calibration offset is affected
by the size of the wavelength bin used in the comparison, 
with a larger bin giving a smaller
formal fit error but less resolution on the wavelength scale over which 
calibration errors occur.  As a compromise we use a bin of 350 $\kms$ or 6\AA\ 
at 5500\AA.  We translate these wavelength calibration shifts into 
velocity using $\vshift = c  \Delta \lambda / \lambda$
and display the result for each exposure in Figures~\ref{fig:vshiftallu}
and \ref{fig:vshiftalll}.

Considering each order from each iodine exposure separately, we began by continuum fitting the exposure and masking the data that fall under strong quasar absorption lines and data near the edges of the order where signal/noise dropped to $< 8$ pixel$^{-1}$. Considering the remaining data as one large wavelength bin, we performed a simultaneous fit for three variables: the overall normalization multiplication factor to the continuum, the Gaussian convolution kernel, and the overall wavelength shift. The wavelength offset found this way is in some sense the average offset for the entire order.
We then held the Gaussian convolution sigma and normalization factor fixed for the whole order and fit for the wavelength calibration shifts in the smaller wavelength bins across the order.  
We fit using a bin size of 350 $\kms$ and report the value of the bin at the average wavelength value 
within the bin. The bins overlap which means the wavelength calibration 
has effectively gone through a smoothing filter of 350 $\kms$.


One sees from Figures~\ref{fig:vshiftallu} and \ref{fig:vshiftalll} that 
there are two distinct types of velocity shifts.  First, there is an exposure-dependent
overall average calibration shift which varies from less than 100 $\ms$ in exposure 2,
to around 500 $\ms$ in exposure 3.  Second, there are intra-order velocity shifts around the overall average
shift ranging in size
from around 100 $\ms$ up to around 400 $\ms$ within each exposure and within each echelle order.
Both types of shifts were also seen in the Keck HIRES iodine cell data (Griest et al. 2010), but
with considerably larger amplitude.
Exposures 1 and 2 have an overall average shift of around 100 $\ms$, 
while exposure 3, taken on the next day is off by about 500 $\ms$.  
It is important to note that the overall average velocity shifts between exposures do not affect 
fine-structure constant work significantly; velocity shifts which vary 
with wavelength -- i.e., effects which would shift one absorption feature with respect to
another at a different wavelength -- are most important (e.g., Webb et al. 1999; Murphy et al. 2009).  
Thus, we are most interested in the relative intra-order shifts within 
each exposure of typically 100 $\ms$ and occasionally as high as 500 $\ms$.
The cause of these intra-order shifts is not completely 
understood, but some possibilities will be discussed below.

\begin{figure}
\begin{center}
\includegraphics[width=0.5\textwidth]{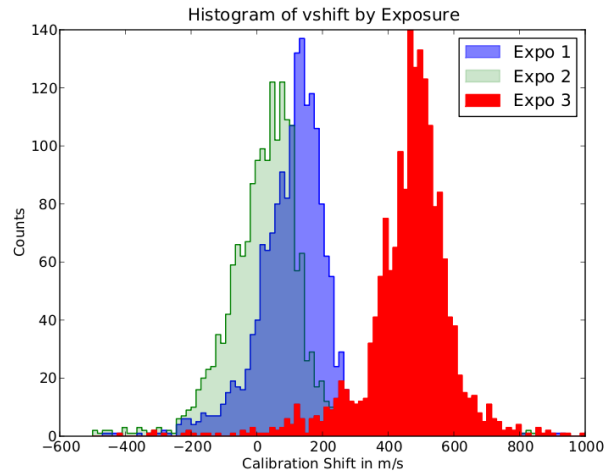}
\caption{Histograms of shifts from the three I$_2$ exposures.
Exposure 1 is plotted in blue (in the middle), exposure 2 in green (on the left)
and exposure 3 in red (on the right). 
}
\label{fig:shifthisto}
\end{center}
\end{figure}

Another way of visualizing the velocity shifts in Figures ~\ref{fig:vshiftallu} and \ref{fig:vshiftalll} is to plot the shift distributions.  These are 
shown in Figure~\ref{fig:shifthisto} which are just
histograms of the shifts for each of the exposures. 
The shift from each fine wavelength bin is given equal weight in
the figure, which, if normalized, can therefore be interpreted as a 
sort of probability of finding a shift of that value.  
We use this histogram in this way below when we investigate the effect these
shifts might have on measurements of $\delalpha$.  Of course, the shift values
for nearby pixels are strongly correlated due to our smoothing, but this
is not important for how we use these distributions below.
We give the values of the means and standard deviations for these histograms in Table~\ref{tab:means}.
Again, it is only the width (sigma) of each histogram that is relevant for fine-structure 
constant work, and not the overall average shift which is the histogram mean.

As discussed above, 
the exposure-dependent, overall average velocity shifts are of the 
magnitude one may expect from errors in the position of the QSO in the 
spectrograph slit.  These types of shifts could also be caused
by grating shifts, temperature/pressure drifts, etc.  
In preliminary work using the UVES iodine cell on exposures toward bright 
stars, we do find that varying the position of the star within the slit can 
give shifts between Th/Ar and iodine
calibrations of around 1500 $\ms$, making this a likely contributor.

Next, we would like to explore the possibility that
there is a relatively constant intra-order distortion that repeats
for each order and is constant between exposures.
If the intra-order distortion pattern for a given object exposure can be well approximated from 
a subsequent I$_2$ exposure, either of the QSO or of a nearby bright star, then one can 
correct the Th/Ar wavelength scale of the former, 
i.e., a sort of calibration transfer function can be established and applied.
We are interested in the effect of wavelength calibration on $\delalpha$,
\begin{equation}
\Delta\alpha/\alpha\propto \Delta v_{ij}/c,
\end{equation}
where $\Delta v_{ij}$ is the velocity difference {\it between} two lines, $i$ and $j$ within the one 
spectrum.

\begin{figure}
\includegraphics[angle=-90,scale=.3]{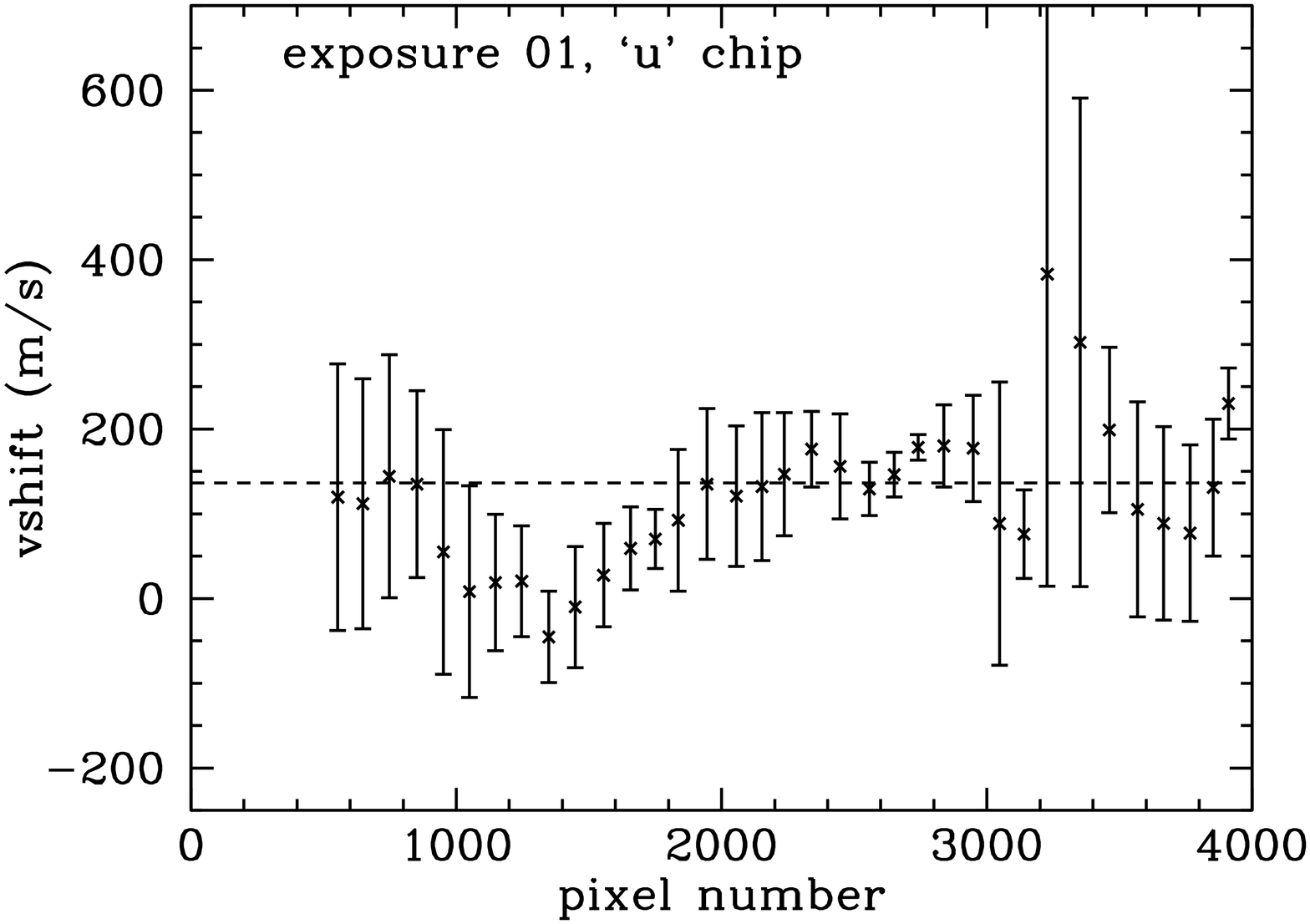}
\includegraphics[angle=-90,scale=.3]{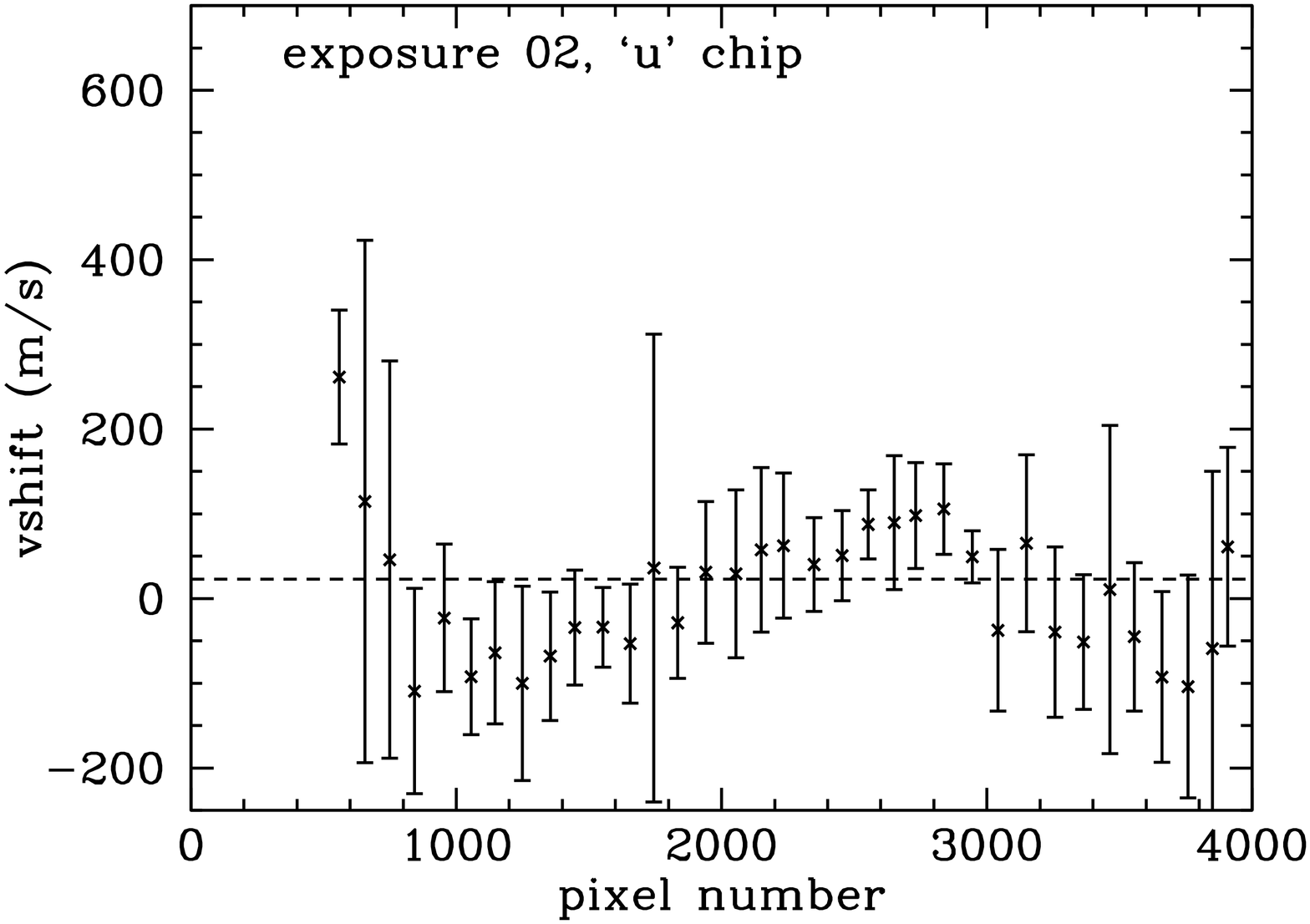}
\includegraphics[angle=-90,scale=.3]{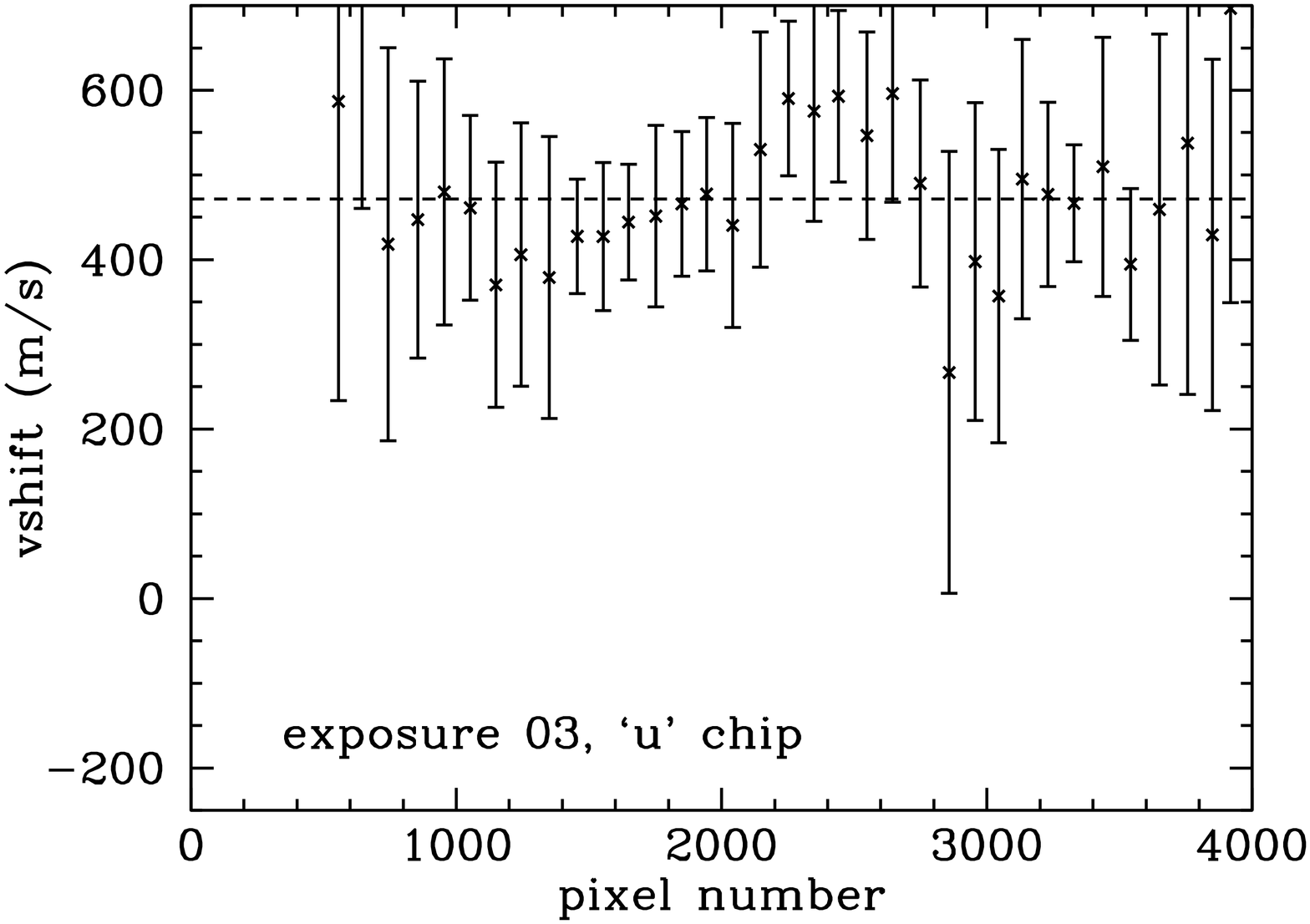}
\caption{
Average over echelle orders of calibration shifts for the ``u'' (upper) chip as a function of pixel number.
The echelle average calibration shift in $\ms$ at
each position bin is plotted on the ordinate.
The standard deviation in the bin of each calibration shift is plotted as the error bar.
The dashed line shows the average over all bins.
\label{fig:orderbinu}
}
\end{figure}

\begin{figure}
\includegraphics[angle=-90,scale=.3]{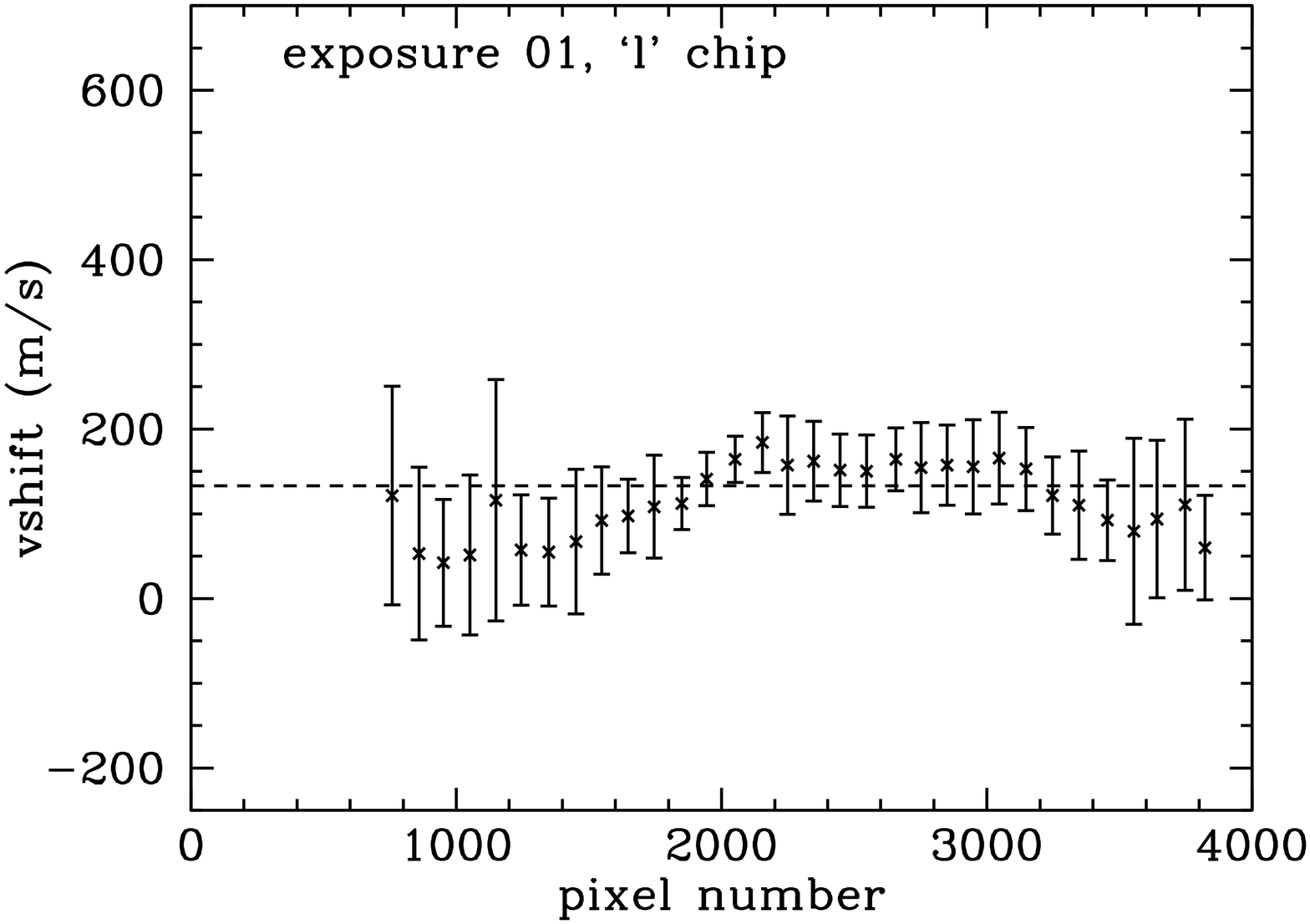}
\includegraphics[angle=-90,scale=.3]{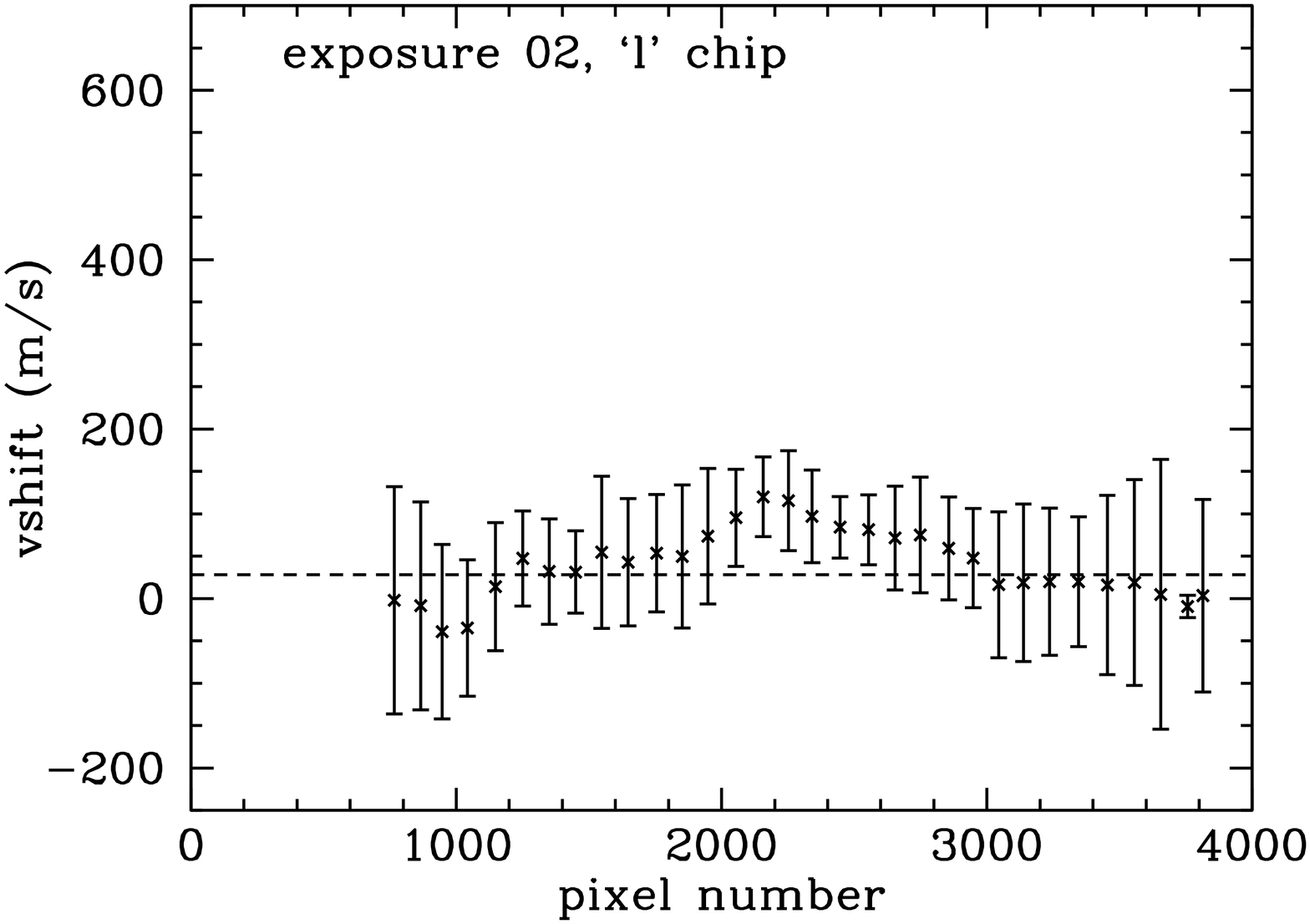}
\includegraphics[angle=-90,scale=.3]{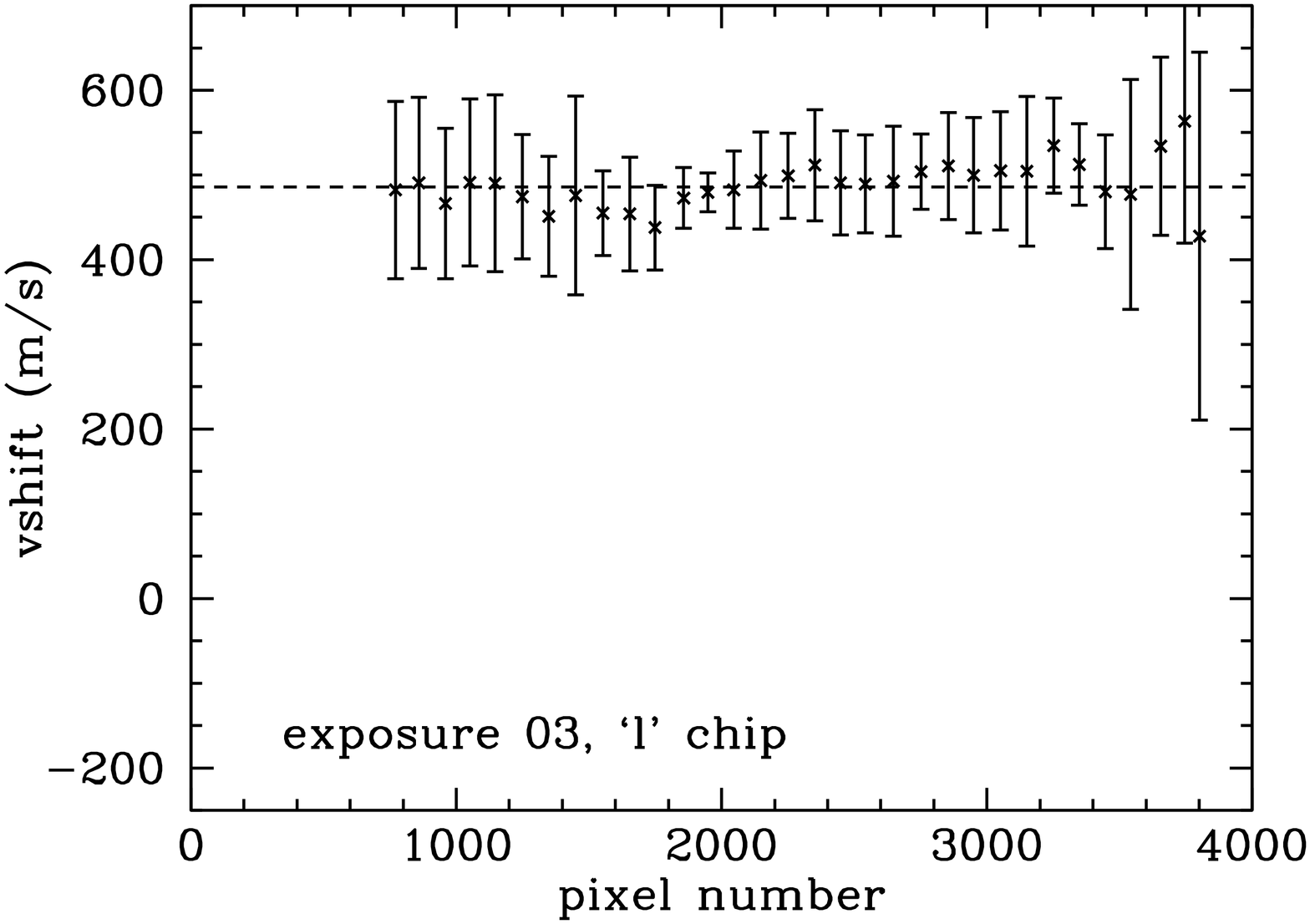}
\caption{
Same as last figure, but for the lower ``l'' chip.
\label{fig:orderbinl}
}
\end{figure}

Examination of Figures~\ref{fig:vshiftallu} and \ref{fig:vshiftalll} might give one
the impression that there is a
pattern of intra-order velocity offsets that repeats in each echelle order.  To test this
hypothesis, we overlaid all the echelle orders of each exposure on their common
CCD pixel scale and then averaged them in bins of 100 pixels. 
We plot the mean and standard
deviation in each bin in Figures~\ref{fig:orderbinu}(a)--(c) and \ref{fig:orderbinl}(a)--(c). 
The dashed lines show the average over all bins.
These figures thus provide the average intra-order distortion for a given exposure.
We do this separately for each exposure since 
the distributions of shifts shown in Figure~\ref{fig:shifthisto} show
large variation between the exposures.  We also treat the UVES upper, ``u,'' chip 
orders separately from the UVES lower, ``l,'' chip orders. 
Thus, we have a total of six figures.

Examination of these figures shows a weak pattern, especially for the ``u'' chip.
We see a roughly linear rise in $\vshift$ of around 200 $\ms$
starting around pixel number 1000 and ending around pixel 2900, where there is a sharp drop of
over 100 $\ms$.  The ``l'' chip figures show smaller overall variance and while exposures 1 and 2 show 
some rise over the first half of the pixels, the amplitude is smaller than for
the ``u'' chip.  Exposure 3 of the ``l'' chip
does not seem to have any pattern of deviation.
Examination of the standard deviations listed in Table~\ref{tab:means} confirms these impressions,
especially that most of the variance comes from the ``u'' chip.  
This is also clear in Figures~\ref{fig:vshiftallu} and \ref{fig:vshiftalll}
where the larger ``u'' chip variances are evident. 

As stated above, 
the purpose of looking for patterns is to find a correction that could be applied to all exposures.
Modeling the  ``u'' chip deviation as a linear rise of 200 $\ms$ over pixels between 1000 and 2900,
we can subtract this line from each of the exposures and recalculate the means
and standard deviations.  The results of this ``correction'' are displayed in Table~\ref{tab:means}.
We only expect this to decrease the standard deviations for the ``u'' chip, but show the ``l'' chip
results also for completeness.  We find a small decrease in standard deviations, especially for
the weighted and sigma-clipped means and standard deviations.  For example, the variance in
exposure 1 ``u'' chip goes from 87  $\ms$ to 73 $\ms$; exposure 2 and exposure
3 also drop their sigma values by around 10 $\ms$.  
We had hoped for more improvement, but with only three iodine exposures, we view this work as
illustrative rather than conclusive.  More data are needed in order 
to further explore the size and constancy of this type of correction.

\begin{deluxetable*}{cccccc}
\tabletypesize{\scriptsize}
\tablecaption{Means$^{\rm a}$ and Standard Deviations$^{\rm a}$ of Calibration Shifts
\label{tab:means}}
\tablewidth{0pt}
\tablehead{
\colhead{Exposure} &
\colhead{Chip} &
\colhead{Unweighted}  & 
\colhead{Unweighted Corrected$^{\rm b}$} &
\colhead{Weighted and Clipped$^{\rm c}$} & 
\colhead{Weighted and Clipped Corrected$^{\rm b}$$^{\rm c}$}}
\startdata
1&l & 118 $\pm$ 76    & 115 $\pm$ 73  & 133 $\pm$ 58  & 123 $\pm$ 52 \\
2&l & 45 $\pm$ 86     & 42 $\pm$ 87   & 63 $\pm$ 69   & 53 $\pm $70 \\
3&l & 487 $\pm$ 75    & 483 $\pm$ 85  & 485 $\pm $60  & 475 $\pm $67 \\
1&u & 116 $\pm$ 138 & 116 $\pm$129 & 107 $\pm$ 87  & 102 $\pm$ 73 \\
2&u & 0 $\pm$ 137   & 0 $\pm$ 130   & -5 $\pm$ 87   & -9 $\pm$ 70 \\
3&u & 477 $\pm$ 174   & 475 $\pm$ 172 & 499 $\pm$ 115 & 495 $\pm$ 105 
\enddata
\tablenotetext{a}{All numbers are in $\ms$}
\tablenotetext{b}{Each pixel has a correction added before calculation. See the text.}
\tablenotetext{c}{The mean is calculated using the error bars as weights and then recalculated after throwing
out points more than 3$\sigma$ from the mean}
\end{deluxetable*}


\subsection{Effect of degree of polynomial in Th/Ar calibration}

\begin{figure}
\includegraphics[angle=-90,width=0.5\textwidth]{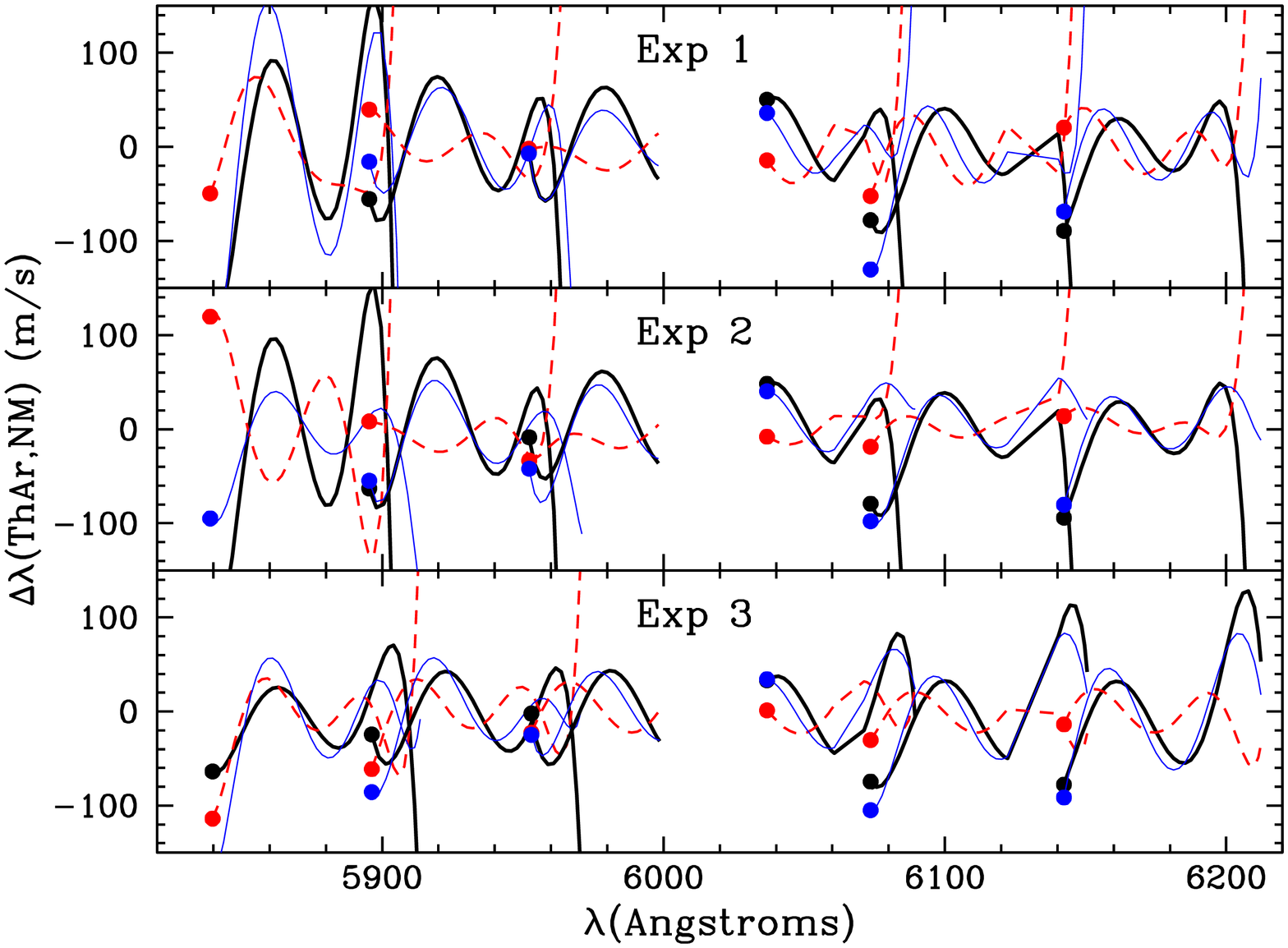}
\includegraphics[angle=-90,width=0.5\textwidth]{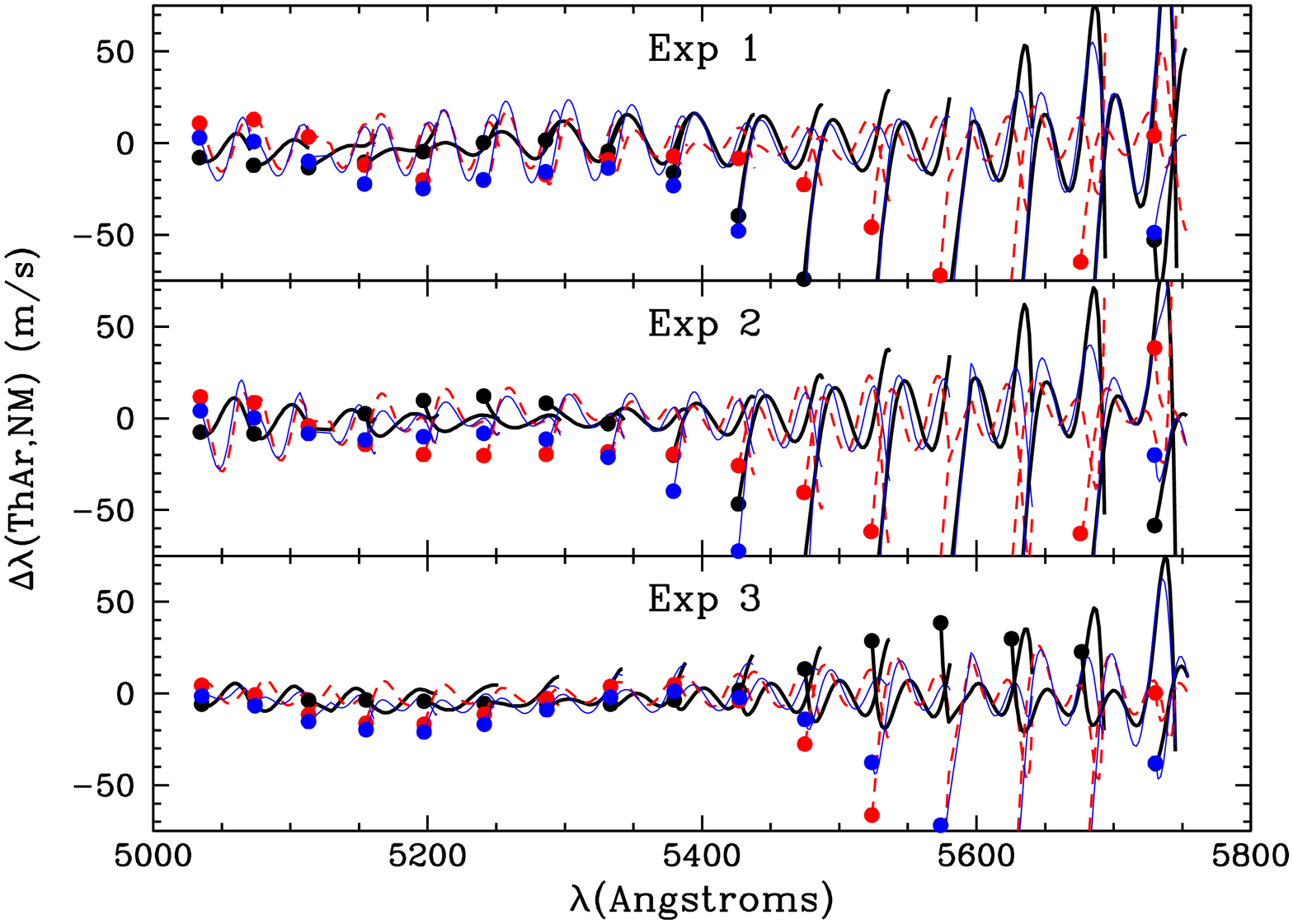}
\caption{
Differences in Th/Ar wavelength solutions for orders occurring on the
upper ``u'' CCD chip (upper
panel) and lower ``l'' CCD chip (lower panel).  
The thick solid black lines show the difference in wavelength scale between
extractions done with a fourth-order polynomial and a fifth-order polynomial.
The thin solid blue lines show the fourth-order minus the sixth-order, and
the dashed red lines show the fifth-order minus the sixth-order.
The beginning of each order is shown by a big dot and exposure number is
labeled.  Notice the different scales on the upper and lower chips.
\label{fig:polythar}
}
\end{figure}

We would like to find the source of the intra-order wavelength 
miscalibrations discovered above, especially whether they come from the 
spectrograph and/or telescope themselves, or from something in the extraction 
and calibration software.
In this section we consider the extent to which the polynomial fit
for the wavelength solution contributes to wavelength scale error.
Thus, besides the standard spectrum extraction using a fourth degree polynomial 
fit for the Th/Ar wavelength solution, we also extracted 
and calibrated the spectra using fifth and sixth degree polynomials.  
Figure~\ref{fig:polythar} shows the differences between these Th/Ar solutions vs.
wavelength.  
The thick black lines shows fourth degree minus fifth degree, 
thin blue lines show fourth minus sixth degree, and dashed red lines
show fifth minus sixth.  The upper panel
shows the upper CCD ``u'' chip, and the lower panel shows the ``l'' chip.
We see that there is substantial disagreement
between these wavelength solutions especially for the
upper CCD and near the edges of the echelle orders.  
For the upper chip, disagreements of 30--40 $\ms$ are typical with
100 $\ms$ not uncommon.  For the lower chip, agreement is typically within
10--20 $\ms$, with outliers mostly near the trailing edge of the orders.
Note that it is not clear that using a sixth-order polynomial does a better
job, since the disagreement between fifth and sixth-order (dashed red line) 
can be appreciable.  

\begin{figure}
\includegraphics[angle=-90,width=0.5\textwidth]{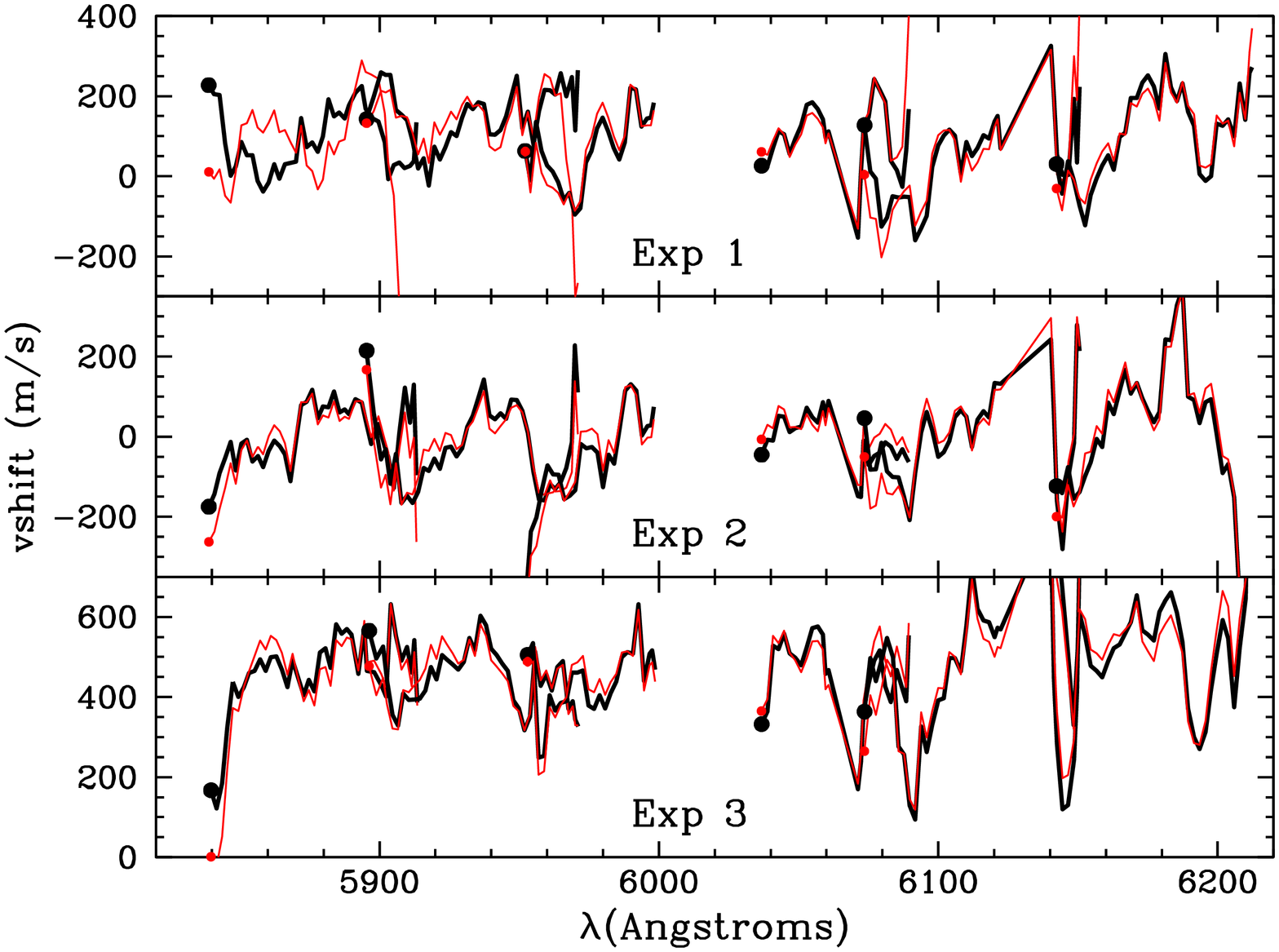}
\includegraphics[angle=-90,width=0.5\textwidth]{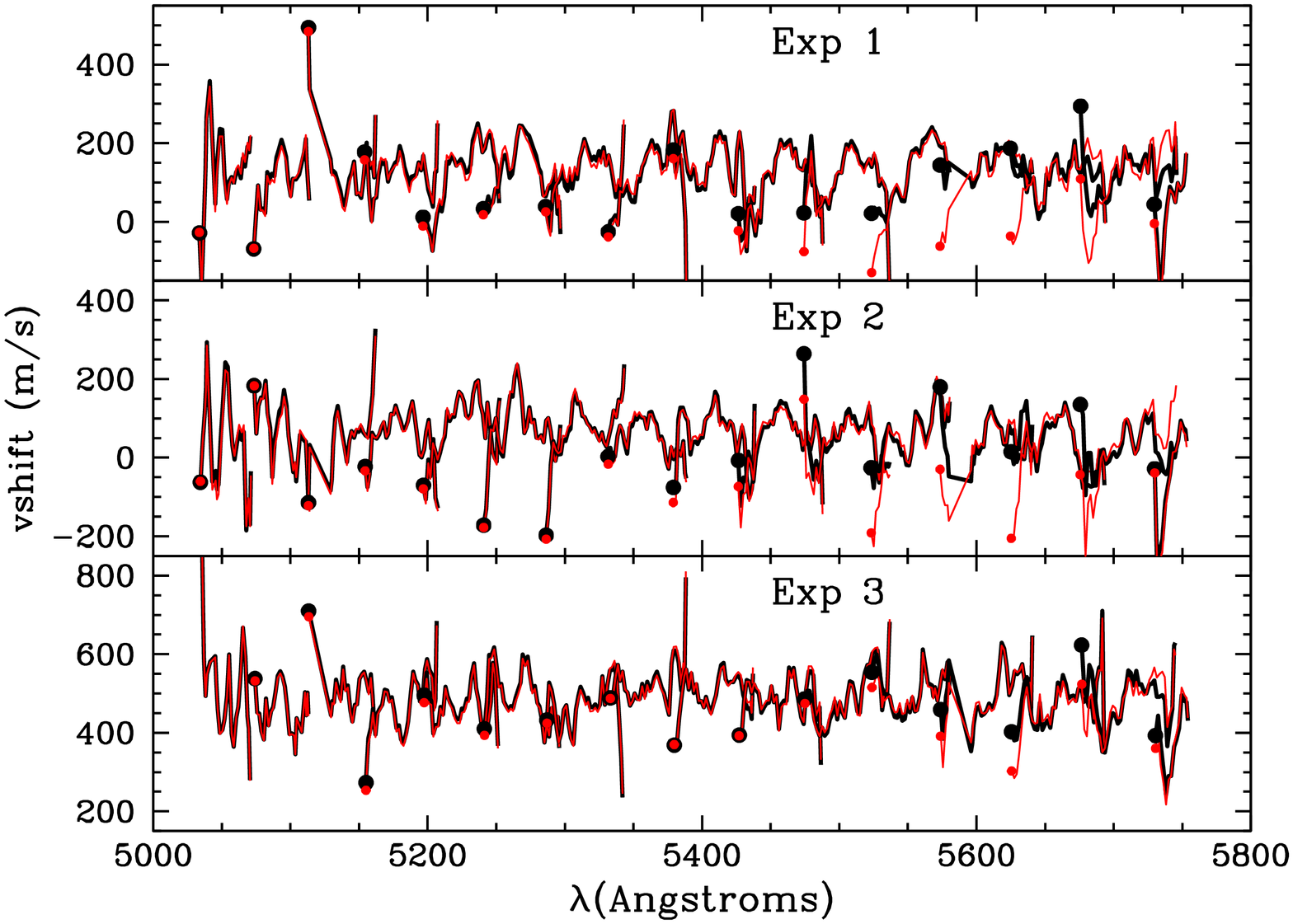}
\caption{
Velocity shift needed to bring the Th/Ar UVES calibration in line with iodine spectrum.
The upper panel shows the results for the three exposures for the upper ``u'' CCD chip, and the lower
panel shows the lower ``l'' chip.
The thick black lines show results for the fourth degree polynomial fit, while the thin red lines show the
sixth degree fit.  The beginning of each echelle order is marked with a large dot.
\label{fig:vshift46}
}
\end{figure}

Next, we recalibrated these spectra using the iodine cell method.  
Figure~\ref{fig:vshift46} shows the velocity shifts needed to bring 
the Th/Ar calibrations
in line with the iodine spectrum.  The black lines are for the fourth degree 
polynomial fit and are the same shifts plotted in Figures~\ref{fig:vshiftallu} 
and \ref{fig:vshiftalll}, 
while the thinner red lines are for the sixth degree polynomial fit.
Note that the distortions in the wavelength scale 
from polynomial fitting (differences between the red and black lines)
are substantially smaller than the total intra-order distortions,
especially for the lower ``l'' chip where the red lines are mostly on top of the black lines.
This shows that polynomial fitting errors are not the primary source of the intra-order distortions.

We also compared the absolute wavelength scales found by the iodine method, 
$\lambdaI(\lambda)$ in Equation~\ref{eqn:io} above, for the fourth, fifth, and sixth degree polynomial fits.
These agreed with each other exceedingly well, with RMS differences between the fourth and sixth degree
fits of 1 $\ms$ or less for the lower ``l'' CCD chip, and RMS differences of 5 $\ms$ or less for
the upper ``u'' chip.  This is a nice confirmation of the robustness of our iodine fitting method,
since this high level of agreement obtained even very near the edges of the echelle orders
where the different polynomial fits disagreed greatly.

Thus, we have several interesting conclusions.
First, we see from Figure~\ref{fig:polythar} that the polynomial extraction algorithm can 
make a significant contribution
to wavelength miscalibration, especially in the upper ``u'' CCD and most
especially near the edges of the echelle orders,
where velocity shifts of order 70 $\ms$ are common. 
Second, there is little evidence that a sixth-order polynomial does a better job
than a fourth-order polynomial, since in both cases the robust iodine cell
recalibration shows similar size miscalibrations.
Thus, we caution against the use of lines near the edges of the echelle orders calibrated only using
Th/Ar.
Third, the overall contribution to the miscalibration variance of the 
polynomial fitting is small compared
to other contributions.  Thus, better treatment of wavelength fitting in
spectra extraction will not cure the main source of intra-order wavelength 
miscalibration we are finding.

\subsection{Possible causes for the intra-order distortions}

While more work is clearly needed for understanding the
ultimate cause(s) of the velocity shifts (both the overall shifts
between exposures and the intra-order distortions within individual
exposures), this will probably require a larger data set with higher
spectral signal/noise and/or spanning a range in several parameters
such as time and telescope pointing. Nevertheless, we briefly
discuss some possible causes for the intra-order distortions between
the iodine and Th/Ar calibrations.

First, for Keck/HIRES, Griest et al. (2010) cited a known
misalignment between the Keck telescope optical axis and that of
HIRES (Suzuki et al.~2003) as a probable cause for the regular shape
of the intra-order distortions from order to order. The misalignment
causes differential vignetting between the telescope beam and that
of HIRES (and, therefore, the Th/Ar beam). While we are unaware of a
similar misalignment in UVES, it remains possible that similar
differential vignetting is responsible for the intra-order
distortions found here. In general, and despite the use of
the simple pupil stop in UVES, the science object and Th/Ar calibration
light do not follow the same optical path, and so we regard a subtle
differential effect of similar, or even subtler nature to be
responsible.

However, other simpler effects may also be suspected at
first. The very similar shape of the intra-order distortions across
many orders in HIRES, and the weaker but broadly similar tendency
revealed in Figure~\ref{fig:orderbinu} and, to a lesser extent, 
Figure~\ref{fig:orderbinl} for UVES, brings
to mind the blaze function of echelle spectrographs. For example,
one may worry that the centroiding of Th/Ar emission lines is
systematically shifted in opposite directions either side of the
order center, thereby producing an intra-order distortion which,
broadly speaking, was similar across all orders. However, this possibility
can be ruled out: the slope produced by the blaze is too small, as
shown by the following simple considerations. In selecting Th/Ar
lines best used for calibrating UVES, Murphy et al.~(2007) measured
the velocity shift between two Gaussian fits to each line, one with
a constant offset continuum, the other with a sloping continuum. It
was found that a background slope of 0.01\,($\kms$)$^{-1}$ in relative
line intensity resulted in a $\sim160$\,ms$^{-1}$ shift in the line
centroid if only a constant offset continuum was fitted. If the
blaze function of UVES is not normalized out of the Th/Ar spectra,
then the background intensity will drop by a factor of 2 over half
the free spectral range (FSR) either side of the center of each
order.  UVES's FSR is $\sim 3000\kms$.  Thus, the average slope of the
inner half of the blaze function is $\sim\pm
0.5/(0.5\times3000\kms)=3\times10^{-4}\,(\kms)^{-1}$ or, since it is
curved, the maximum slope will certainly be
$\lsim$$1\times10^{-3}$. This would translate to a line shift of
only $160{\rm \,ms}^{-1} \times 1\times10^{-3}/0.01 = 16\,{\rm
ms}^{-1}$. This upper limit is an order of magnitude below the
intra-order distortions we observe.

The above blaze effect will not appear in our UVES Th/Ar
calibration because the Gaussian fits to the Th/Ar lines had a sloped
continuum. Also, our Th/Ar extractions involve a crude blaze
correction; even if this did not reliably remove all blaze effects,
the resulting line-shifts would be smaller than the above
calculation indicates. For example, the very broad spectral shape of
the quartz lamp which provides the flat field light will remain in
our CPL-reduced Th/Ar spectra (and the I$_2$ spectra), but the above
argument demonstrates that such a broad background spectral shape
will have negligible effects and cannot explain the intra-order
distortions we find. 
Similar blaze effects probably play little role
in the HIRES results of Griest et al. (2010) because the data
reduction pipeline which provided their main results 
(HIRES REDUX by Jason X.~Prochaska) uses a
sloped continuum.  They also used another pipeline (MAKEE by Tom
Barlow) that used a constant offset continuum, but found similar results.
That is, any effect from
the blaze function in the Griest et al.~results will be small by
comparison to the intra-order distortions measured there.

Recently, Wilken et al.~(2010) compared
the wavelength scales established from a laser frequency comb (LFC)
and a standard Th/Ar lamp in one echelle order of the HARPS
spectrograph on the ESO 3.6 m telescope. The LFC revealed periodic
artifacts in the physical pixel size and/or sensitivity from the CCD
manufacturing process. By appropriately correcting the sensitivity
of every 512th pixel, the wavelength calibration residuals were
substantially improved. They also found distortions between this LFC
calibration and that from the Th/Ar lamp. Perhaps the CCD
manufacturing defects cause this distortion and contribute to the
intra-order distortions we find for UVES. This seems unlikely
because the Th/Ar lines are distributed randomly and very sparsely
across the CCD grid, so very few (if any) Th/Ar lines straddle a
pixel with a manufacturing defect. In fact, the Th/Ar--LFC
distortions found by Wilken et al.~seem similar in shape and
magnitude to those we find for UVES between different Th/Ar solutions
using different polynomial degrees; compare Figure~\ref{fig:polythar} with their Figure
4 (blue line).

Finally, we note that a similar method for discovering shifts
in the Th/Ar calibration of UVES has been demonstrated by Molaro et
al.~(2008b).  They compared the solar spectrum
reflected from asteroids with a laboratory FTS solar spectrum to
establish the wavelength scale of UVES and compare it with the Th/Ar
solution. First, it is interesting to note that they only find
small (10--50\,m\,s$^{-1}$) average velocity shifts for different
exposures, possibly indicating that accurate centroiding of the
asteroids within the 0.5'' wide slit was achieved. However,
only the line cores of a few solar absorption features per echelle
order were utilized, so it is difficult to assess whether the
intra-order distortions we find here are reflected in the results of
Molaro et al.~(2008b).

\section{Effect on $\delalpha$}

As stated in the introduction, wavelength calibration errors of the magnitude
presented in this paper are unimportant for most astronomical work.
In this section we wish to address the question of whether or not calibration
errors such as we are finding can make a difference in experiments looking for
changes in $\delalpha$.  In order to proceed we need an estimate of the magnitude
of $\delalpha$ we wish to be sensitive to.  There are at least two possible 
sources for this number:  
first, $\delalpha = (-5.7 \pm 1.1)\ten{-6}$, which is the most robust claimed 
detection of $\delalpha$ (Murphy et al. 2001a, 2001b, 2003, 2004), and second there are several claimed limits on $\delalpha$ near 
$\delalpha < 1 \ten{-6}$ (Chand et al. 2004; Srianand et al. 2004;
Levshakov et al. 2006, 2007; Molaro et al. 2008a; cf. Murphy et al. 2008).
Thus, the question we would like to ask 
is to see whether the UVES wavelength calibration errors discussed above,
can alone give rise
to estimates of $\delalpha$ in the $1\ten{-6}$ to $5\ten{-6}$ range.

Naively, one might expect the calibration errors of 200 $\ms$ to 500 $\ms$ we found
above to give rise to spurious detections of $\delalpha$ of order
$1\ten{-6}$.  If $\alpha$ was different
in the past, the velocity shift of an absorption line from the lab value due
to changing $\alpha$ can be found from
\begin{equation}
v_j  = v_0 +\left(\delalpha\right) x_j,\qquad\qquad  x_j  = -2 c  q_j \lambda_{0j},
\label{eqn:eqvj}
\end{equation}
where $j$ numbers the atomic transitions that are being compared,
$v_0$ is a constant offset (degenerate with the system redshift), the $x_j$ are
constants that depend only on the wavelength $\lambda_0$ of the transition,
and the $q_j$ characterize the sensitivity of each transition to 
a change in $\delalpha$.
For transitions of interest in QSO absorption work, these $q$-values range
in magnitude from around 20 ${\rm cm}^{-1}$ to 2000 ${\rm cm}^{-1}$
(Porsev et al. 2007), giving rise to
velocity shifts of a few meters per second up to around 100 $\ms$ (for a value of 
$\delalpha=-5.7\ten{-6}$).
Thus, for individual transitions, the sizes of the wavelength
calibration errors found above are of the order of or larger than the sizes of 
the signal expected from a changing $\alpha$.

However, the value of $\delalpha$ obtained does not depend upon just one transition,
but is always a comparison of two of more transitions.  Thus, an overall velocity
shift of 500 $\ms$ as seen in exposure 3 above is not relevant as long as comparisons
are not done across different exposures.  Figures~\ref{fig:vshiftallu} 
and \ref{fig:vshiftalll} show that
the relative shift across one exposure is substantially smaller than
the overall shift.  In addition,
if the calibration shift error is random, and equally likely to 
be negative as positive, then it can be removed by averaging over many transitions
and many lines of sight.  Potential problems do exist if the distribution of
relative shifts is not random, or 
if the sizes of the errors are large enough that
a good measurement requires too many transitions or too many lines of sight.
Thus we wish to perform some mock experiments using the above distributions
of calibration shifts to see what the effect on $\delalpha$ would be.

In an actual experiment, a value of $\delalpha$ is estimated by doing a
large joint fit of the Voigt line profiles, the system redshifts, 
and the possible velocity offset due to $\delalpha$.
In our Monte Carlo mock experiments we try to calculate a value of $\delalpha$ 
using the UVES VLT data, but without measuring any actual absorption lines.  
Instead of using the fitted system redshifts to find velocity offsets, we use
the wavelength calibration offsets given by the fine-binned (blue and red) 
lines in Figures~\ref{fig:vshiftallu} and \ref{fig:vshiftalll}, and add
these to the lab values of $\lambda_0$ in Equation~(\ref{eqn:eqvj}).

The basic method is as follows.
We first choose a random redshift in the range $z=0.2$ to $z=3.7$, and calculate
the wavelengths of the 23 atomic transitions that were
studied in Murphy et al. (2003).  
We define $\Ntran$, the number of these 23 transitions that 
fall at wavelengths
for which we have iodine wavelength calibration.  We require at least $\Nmin$
transitions, and show our results as a function of this $\Nmin$ (a typical value
is $\Nmin=4$, and we do not find any cases with $\Ntran>9$).  
For each such transition we shift its wavelength by an amount
given by the 
fine-binned (blue or red) line from one of the exposures in 
Figures~\ref{fig:vshiftallu} and \ref{fig:vshiftalll}.  We then
perform a fit of Equation~(\ref{eqn:eqvj}) for $\delalpha$ and its error.
This counts as one absorption system, and we repeat this procedure $\Nsys$ 
times, averaging the values of $\delalpha$ obtained.
We consider values of $\Nsys$ ranging from $\Nsys=143$, the number of systems 
used in Murphy et al. (2004), to
$\Nsys=1$, the value when only one system in one QSO is being analyzed.
The above procedure constitutes one Monte Carlo experiment.  We repeat the 
experiment many times to find an average value of $\delalpha$ and its standard
deviation (measured by the variance of $\delalpha$ for the many experiments).

We present resulting average values of $\delalpha$ along with their 
standard deviations as a function of $\Nsys$ and $\Nmin$
in Table~\ref{tab:monteresults} for 200,000 mock experiments.  
The table shows results for $\Nsys=143$ and $\Nsys=1$.  Note one expects
the standard deviation in $\delalpha$ to simply scale as $\Nsys^{-1/2}$,
which is close to what we find in Table~\ref{tab:monteresults}.  Thus, we will use
this scaling from now on and only report Monte Carlo experiments for $\Nsys=1$.

Besides using the actual wavelength calibration errors above, we also 
ran several Monte Carlo simulations using two simple models of 
the calibration offsets.
The results of these simple models are also reported in
Table~\ref{tab:monteresults}.  
For the first model we used 
a Gaussian random velocity offset with a standard deviation equal to 91 $\ms$.
For the second, we modeled the velocity offsets as
a sine function with amplitude equal to $\pi/2$ times
the standard deviation of the velocity offsets for one of the exposures above,
and with a wavelength of about one echelle order.
For this case, we found the results did not depend strongly on 
the sine wavelength. 
As seen in Table~\ref{tab:monteresults} the results for $\sigma(\delalpha)$ are
quite similar for all three exposures and for the Gaussian and sine function
models.

Table~\ref{tab:monteresults} shows several interesting things.
First in all cases when $\Nmin=2$, both the mean calibration offset and
standard deviation in $\delalpha$ are substantially larger than expected from a simple
$1/\sqrt{\Ntran}$ scaling.  We think this is due to occasional cases where there
are very few transitions found, but these lie close together in 
$x_j= -2cq_j \lambda_j$.
Since $\delalpha$ is basically the slope in Equation~(\ref{eqn:eqvj}), a small
$\Delta x$ offset can result in a very large slope and 
therefore a large error in 
$\delalpha$.  It takes just a few such
cases to greatly increase the standard deviation.  
A lesson here may be not to use systems in which very few transitions can
be compared.  For example, in the alkali doublet method 
(e.g., Bahcall et al. 1967, Varshalovich et al. 2000, etc.)
two transitions that are close together in wavelength are compared, 
so this method would be sensitive to intra-order distortions.
More generally, even for $\Nmin=4$ and $\Nmin=6$, we find the errors dropping
more quickly with $N$ than $1/\sqrt{N}$.

Next, we note that the results for all three exposures and for the Gaussian
and sine function error models are quite consistent,
especially when one takes into account
that the standard deviation of velocity offsets for exposure 1 is slightly smaller
than for the other exposures.  Also as expected, the large overall velocity
shift for exposure 3 had no effect.

If we restrict ourselves to the $\Nmin=6$ column, and consider $\Nsys=143$,
we see that the systematic error introduced to a many multiplet
measurement of $\delalpha$ is around $0.28 \ten{-6}$,
significantly smaller than the statistical error of $1.16 \ten{-6}$  stated in
Murphy et al. (2003, 2004).  We do note that Murphy et al. (2003, 2004) 
used the Keck HIRES spectrograph and not the VLT-UVES instrument.

An important problem with the above results is that the value of $\delalpha$ and
its standard deviation depends strongly on $\Ntran$, the number of transitions
compared in each system, and that no cases were found with
$\Ntran>9$.  This latter fact is because the results above only 
included lines that 
overlapped with our iodine cell coverage. 
Thus, $\Ntran$ found in our Monte Carlo experiments are 
artificially lower than in an actual experiment, which typically has more
spectral coverage.  Since the value of $\sigma(\delalpha)$ drops quickly
with an increase $\Ntran$, we also expect an actual
experiment to find smaller deviations than the ones we report.
We find this low value of $\Ntran$ to be especially true for certain values 
of $z$, where very few interesting
lines fall within our iodine cell coverage.
Our attempt to get around this by setting a minimum number
of transitions, $\Nmin$, was partially successful, but 
Table~\ref{tab:monteresults} shows strong dependence on $\Nmin$.  
Even more clearly, we see this in Table~\ref{tab:ntran} where the values of sigma 
of $\delalpha$
depend on $\Ntran$ substantially more strongly than $1/\sqrt{\Ntran}$.
We note that when $\Ntran$ is specified in the table, exactly $\Ntran$ 
transitions are compared, while when $\Nmin$ is specified all cases that
have $\Ntran\geq\Nmin$ are used.

To remedy this situation, we need to somehow estimate the calibration offsets
in regions of the spectra where we do not have iodine cell coverage.
We attempt to do this by replicating the calibration offsets from
the regions where we measure them to all the other spectral regions
where interesting transitions occur; i.e., we assume that
the distributions of shifts illustrated in Figures~\ref{fig:vshiftallu}
and \ref{fig:vshiftalll} apply to all wavelengths.  
We then repeat the Monte Carlo experiments above.  

\begin{figure}
\includegraphics[width=0.5\textwidth, angle=0]{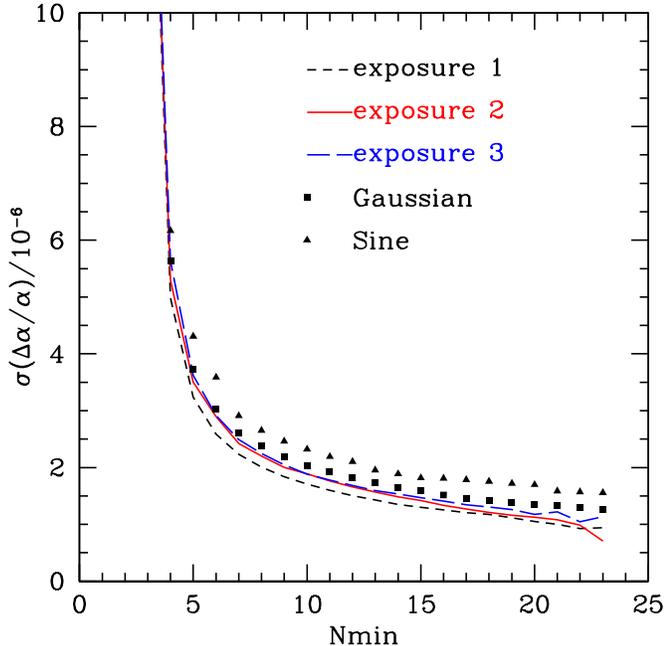}
\caption{Scaling of $\sigma(\delalpha)$ with $\Nmin$, the minimum number 
of transitions allowed in a system.  This is for full wavelength coverage
between 3000 \AA\ and 10,500 \AA, with repeated calibration errors.
200,000 Monte Carlo realizations were used with $\Nsys=1$.
\label{fig:sigvsnmin}
}
\end{figure}

Results of these simulations are shown
in Figure~\ref{fig:sigvsnmin}
and in Table~\ref{tab:monteresultsrep}. 
Now we find few systems with less than 10 transitions and many with around
18. 
The values of $\sigma(\delalpha)$ 
for low values $\Nmin$ are similar to those found in the iodine
coverage only Monte Carlos, again depending very strongly on $\Nmin$ (or $\Ntran$).
However, for larger values of $\Nmin$, $\sigma(\delalpha)$  
stabilizes and approaches the expected $1/\sqrt{\Ntran}$ behavior as 
$\Nmin$ increases.

The strong dependence of $\sigma(\delalpha)$ on $\Nmin$ is interesting and 
suggests that the standard lore 
that says it is better to use transitions in the same echelle order may not be true.  
The wavelength calibration errors
we found above exist even within single echelle orders. 
Thus, it may be more robust to
measure or limit $\delalpha$ using transitions that are well separated in $x$;
the extra ``lever arm'' may be advantageous in reducing scatter.

Overall, the large $\sigma$'s found at small $\Ntran$ imply
that due to wavelength calibration errors it may be dangerous to attempt to
measure $\delalpha$ using only a few transitions in one or two systems.
Until better understanding is found as to the source of these calibration
errors it is probably important to average the errors away by using many transitions
in many absorption systems.

We also see from Figure~\ref{fig:sigvsnmin} that the mean and standard deviation
for all three exposures agree well, and that the three experimentally determined
measurements of $\sigma(\delalpha)$ are well modeled by a Gaussian (or sine 
function) model with the same velocity standard deviation.  Thus, it seems that the
error in $\delalpha$ caused by this wavelength calibration is completely
specified by just the standard deviation of the velocity offsets.
In order to quantify this we ran a suite of simulations using various values
of $\sigma(v)$.  We found we could fit all the results for $\sigma(\delalpha)$ 
with a fairly simple formula:
\begin{equation}
\sigma(\delalpha) =  7.5 \ten{-8} \CNt {\sigma(v) \over (\Nsys \Ntran)^{1\over2}},
\label{eqn:sigmada}
\end{equation}
where $\sigma(v)$ is in $\ms$, and $\CNt \approx 1$ (to within 10\%) for $\Ntran \geq 8$, and 
$\CNt=21$ for $\Ntran=2$, $\CNt=5.1$ for $\Ntran=3$, $\CNt=1.7$ for $\Ntran=4$, 
$\CNt=1.3$ for $\Ntran=5$, and $\CNt=1.1$ for $\Ntran=6$ or $\Ntran=7$.
The simulations were done with our ansatz for full wavelength calibration between 3000\AA\
and 10,500\AA.

\section{Conclusions}
We tested the accuracy of the wavelength calibration of the VLT UVES spectrograph by recalibrating 
the standard Th/Ar calibration pipeline output using spectra taken though the UVES iodine cell.
We found several types of miscalibration:  first, an exposure-dependent overall average velocity shift
that is probably mostly due to the position of the QSO within the spectrograph slit. 
Repositioning of the spectrograph gratings and pressure/temperature drifts inside the spectrograph
may also contribute to these velocity shifts.  These overall average shifts range up to 500--600 $\ms$, but
do not affect fine-structure constant measurements as long as comparisons between absorption features
are only performed within the same spectrum.

Second, and more importantly, we found intra-order calibration shifts of up to several hundred meters/second
that occur within the same spectrum and within each echelle order.  These can affect measurement
of the fine-structure constant.
We investigated several possible causes for these shifts.  We explored the effect of the degree of
the polynomial used in the Th/Ar calibration and found a velocity dispersion of 
20--40 $\ms$ due to this.  
This is several times smaller than the total dispersion we find in the velocity measurements, 
so the main effect we are finding is not due to the degree of the 
Th/Ar calibration polynomial.  
In fact, we found that the iodine
cell recalibration of extractions using different degree polynomials allowed us to regain the same 
absolute wavelength scale to an accuracy of better than 5 $\ms$, thus showing the robustness of
our iodine recalibration method.  We considered several other possible causes for 
the intra-order shifts and showed that spectrograph blaze function
and related possibilities were unlikely.  
Thus, we conclude that most likely there are unknown hardware-related systematic
errors within UVES and/or the VLT responsible for these shifts.   
We noticed some weak patterns in the shift as a function of CCD pixel and 
attempted a correction based upon these patterns.  This correction
was not very successful, so we conclude that additional data and analysis are required to discover
the cause and potential fixes for these wavelength scale shifts.
We note that it is important look for such a correction, since
if the intra-order distortion pattern for a given object exposure could be well approximated from 
a subsequent I$_2$ exposure, either of the QSO or of a nearby bright star, then one could
correct the Th/Ar wavelength scale of the former, 
i.e., a sort of calibration transfer function could be established and applied.

We next explored the effect of these shifts on determination of or limits on the value of the fine-structure
constant.  Using our measured shifts we performed Monte Carlo experiments and found 
that the effect of the intra-order shifts is well modeled 
by a Gaussian dispersion in wavelength
calibration with a magnitude of around 80--120 $\ms$.  We found that the effect on $\delalpha$ depended
strongly on the number of lines being compared within the spectrum, with the use of a small number
of lines resulting in larger than expected errors.  Thus, until a correction to these intra-order wavelength
miscalibrations is found, we recommend against focusing on comparison between pairs of lines or the use 
of spectra which contain only a few lines. We summarized the results of our Monte Carlo experiments in a fitting formula Equation~(\ref{eqn:sigmada}),
which should be useful in estimating the effect of these (or similar) wavelength calibration problems
on future fine-structure work.

\acknowledgements
J.B.W. and K.G. were supported in part by the U.S. Department of Energy
under grant DE-FG03-97ER40546. M.T.M. thanks the Australian Research Council 
for a QEII Research Fellowship (DP0877998).

\begin{deluxetable}{ccccc}
\tablecaption{Monte Carlo Results$^{\rm a}$ for Mean and Standard Deviation 
of $\delalpha$
\label{tab:monteresults}}
\tablewidth{0pt}
\tablehead{
\colhead{Exposure} &
\colhead{$\Nmin$} &
\colhead{$\delalpha$ Mean/$10^{-6}$} &
\colhead{$\sigma/10^{-6}$ for $\Nsys=1$} &
\colhead{$\sigma/10^{-6}$ for $\Nsys=143$}}
\startdata
1 & 2 & 1.74 & 41.6& 3.57\\
1& 4 & 0.53 & 5.21& 0.443\\ 
1& 6  & -0.041 & 3.28& 0.267\\
\tableline
2& 2 &  -1.68&  44.7& 3.74 \\
2& 4 &  0.330&   6.41& 0.540\\
2& 6 &  0.297&  3.36& 0.279\\
\tableline
3& 2 &  1.02&  58.9& 4.83\\
3& 4 &  0.706&  6.67& 0.548\\
3& 6 &  0.245&  3.43&  0.280\\
\tableline
Gaussian & 2 &  -0.502&  115 &$\cdots$\\
Gaussian & 4 &  -0.080&  11.2 &$\cdots$\\
Gaussian & 6 &  -0.012&  4.23 &$\cdots$\\
\tableline
Sine & 2 &  -1.78&  104& $\cdots$\\
Sine & 4 &  -3.13&  12.6& $\cdots$\\
Sine & 6 &  -0.308&  4.64& $\cdots$
\enddata
\tablenotetext{a}{Only transitions within the iodine cell coverage are included
from 200,000 realizations of $\Nsys=1$.}
\end{deluxetable}

\begin{deluxetable}{cccc}
\tablecaption{Standard Deviation$^{\rm a}$ of $\delalpha$ from Wavelength Calibration
Errors as a Function of the Number of Transitions.
\label{tab:ntran}}
\tablehead{
\colhead{$\Ntran$} &
\colhead{Percentage of Realizations} &
\colhead{Exposure 1: $\sigma/10^{-6}$} &
\colhead{Gaussian: $\sigma/10^{-6}$}} 
\startdata
2 & 20.7\% & 228 & 91.4 \\
3 & 29.7\% & 30.3 & 44.0 \\
4 & 18.7\% & 8.61 & 11.2 \\
5 & 11.7\% & 5.90 & 5.00 \\
6 & 10.7\% & 3.34 & 4.17 \\
7 & 6.3\% & 3.10 & 3.73\\
8 & 1.3\% & 3.36 & 3.60 \\
9 & 1.0\% & 2.59 & 3.18 
\enddata
\tablenotetext{a}{For transitions that occur within the iodine cell coverage
region of the spectrum, and for 200,000 realizations of $\Nsys=1$}
\end{deluxetable}

\begin{deluxetable}{ccccc}
\tablecaption{Monte Carlo Results$^{\rm a}$ for Mean and Standard Deviation 
of $\delalpha$
\label{tab:monteresultsrep}}
\tablewidth{0pt}
\tablehead{
\colhead{$\Nmin$} &
\colhead{Exposure 1: Mean/$10^{-6}$} &
\colhead{Exposure 1: $\sigma/10^{-6}$} &
\colhead{Gaussian: Mean/$10^{-6}$} &
\colhead{Gaussian: $\sigma/10^{-6}$}} 
\startdata
2 & -0.018 & 81.2 & 0.185 	& 91.6 \\
4 & -0.101 & 4.99 & -0.036	& 5.64 \\
6 & -0.064 & 2.59 & -0.028	& 3.04 \\
8 & -0.051 & 2.00 & -0.006	& 2.38 \\
10& -0.042 & 1.71 & -0.005      & 2.03 \\
15& -0.029 & 1.30 & 0.004	& 1.59 \\
20& 0.153  & 1.05 & 0.007	& 1.35 
\enddata
\tablenotetext{a}{All of the 23 transitions falling between 
3000\AA\ and 10500\AA\ are included 
(see the text), for 200,000 realizations of $\Nsys=1$.}
\end{deluxetable}


\begin{references}
\reference{} Bahcall, J. N., Sargent, W. L. W., \& Schmidt, M. 1967, ApJ, 149, L11
\reference{} Butler, R. P., Marcy, G. W., Williams, E., McCarthy, C., Dosanjh, P., \& Vogt, S. 1996, PASP, 108, 500
\reference{} Chand, H., Srianand, R., Petitjean, P., \& Aracil, B. 2004, A\&A,
	417, 853 %
\reference{} D'Odorico, S., et al., 2000, Proc. SPIE, 4005, 121
\reference{} Griest, K., Whitmore, J. B., Wolfe, A. M., Prochaska, J. X., Howk, J. C., \& Marcy, G. W. 2010, ApJ, 708, 158
%
%
%
 %
%
%
\reference{} Johnson, J. A., et al. 2006, ApJ, 647, 600
\reference{} Konacki, M., 2009, in Transiting Planets, ed. Pont, F., Sasselov, D. D., \& Holman, M. J. (Cambridge: Cambridge Univ. Press), 141
\reference{} Levshakov, S. A., Molaro P., Lopez, S., D'Odorico, S., Centurión M., Bonifacio, P., Agafonova, I. I., \& Reimers, D. 2007, A\&A, 466, 1077
 %
\reference{} Levshakov, S. A., Centurión M., Molaro P., D'Odorico, S., Reimers, D., Quast, R., \& Pollmann, M. 2006, A\&A, 449, 879
 %
%
\reference{} Molaro, P., Reimers, D., Agafonova, I. I., \& Levshakov, S. A.,
2008a, Eur. Phys. J. Spec. Top., 163, 173 %
\reference{} Molaro, P. et al. , 2008b, A\&A, 481, 559
\reference{} Murphy, M. T., Webb, J. K., Flambaum, V. V., Dzuba, V. A., Churchill, C. W., Prochaska, J. X., Barrow, J. D., \& Wolfe, A. M. 2001a, MNRAS, 327, 1208 
 %
\reference{} Murphy, M.~T., Webb, J.~K., Flaumbaum, V.~V., Churchill, C.~W.,
Prochaska, J.~X., 2001b, MNRAS, 327, 1236 
 %
\reference{} Murphy, M.~T., Webb, J.~K., \& Flambaum, V.~V., 2003, 
	MNRAS, 345, 609 %
\reference{} Murphy, M.~T., Webb, J.~K., \& Flambaum, V.~V., 2004,
	VizieR Online Data Catalog, 734, 50609 %
%
%
\reference{} Murphy, M.~T., Tzanavaris, P.~T, Webb, J.~K., \& Lovis, C., 2007, 
	MNRAS, 378, 221 %
\reference{} Murphy, M.~T., Webb, J.~K., \& Flambaum, V.~V., 2008, 
	MNRAS, 384, 1053 %
\reference{} Murphy, M.~T., Webb, J.~K., \& Flambaum, V.~V., 2009,	Mem. Soc. Astron. Ital., 80, 833
\reference{} Osterbrock, D. E., Waters, R. T., Barlow, T. A., Slanger, T. G., \& Cosby, P. C. 2000, PASP, 112, 733 %
\reference{} Porsev, S. G., Koshelev, K. V., Tupitsyn, I. I., Kozlov, M. G., Reimers, D., \& Levshakov S. A. 2007, Phys. Rev. A 76, 052507 %
\reference{} Srianand, R., Chand, H., Petitjean, \& P., Aracil, B. 2004,
	Phys. Rev. Lett., 92, 121302 %
\reference{} Suzuki, N., Tytler, D., Kirkman, D., O'Meara, J. M., \& Lubin, D. 2003, PASP, 115, 1050
\reference{} Varshalovich, D. A., Potekhin, A. Y., Ivanchik, A. V., 2000, in AIP Conf. 506, X-Ray and Inner-Shell Processes, ed. R. W. Dunford et al. (Argonne, IL: Argonne National Laboratory)
\reference{} Webb, J. K., Flambaum, V. V., Churchill, C. W., Drinkwater, M. J., \& Barrow, J. D. 1999, Phys. Rev. Lett., 82, 884
\reference{} Wilken, T., et al., 2010, MNRAS, 405, L16


\end{references}
\end{document}